\documentclass[
  reprint,
  superscriptaddress,
  amsmath,amssymb,
  aps,
  prc, 
  floatfix,
  nofootinbib
]{revtex4-2}

\usepackage{float}
\usepackage{graphicx}
\usepackage{braket}
\usepackage{xcolor}
\usepackage{placeins}
\usepackage[colorlinks=true,
            linkcolor=blue,
            citecolor=blue,
            urlcolor=blue,
            hyperfootnotes=false]{hyperref}
\usepackage{multirow}

\makeatletter
\def\@hangfrom@section#1#2#3{#1#2#3}
\renewcommand\thesection{\arabic{section}}
\def\sec@upcase#1{#1}

\makeatother

\makeatletter
\renewcommand\subsection{%
  \@startsection{subsection}{2}{\z@}%
  {-2.5ex \@plus -1ex \@minus -.2ex}%
  {1.0ex \@plus .2ex}%
   {\centering\normalfont\itshape}%
}
\makeatother

\begin{document}

\title{ \boldmath{
Critical coupling with zero-mode corrections in discretized light-cone quantized $\phi^4$ theory}}
\date{\today}



\begin{abstract}
We present an advancement for solving the Discretized Light-Cone Quantization (DLCQ) Hamiltonian for mass spectra in 2D $\phi^4$ theory that incorporates perturbative zero-mode contributions. 
We demonstrate that this correction accelerates numerical convergence of the mass eigenstates with increasing resolution and yields a critical coupling of comparable accuracy with a substantial reduction in computational costs. 
Specifically, with more than one order of magnitude reduction in basis space dimensionality  
we achieve an extrapolated critical coupling $\lambda_{c}/\mu^{2}$ of $23.10 \pm 0.25$ compared with $23.53 \pm 0.26$ in the larger basis but without zero-mode correction.  We employ Gaussian Process Regression for extrapolation to the continuum limit.  The approach we present here is prospective for studies of higher-dimensional gauge theories.
\end{abstract}
\author{Shreeram Jawadekar}
\affiliation{Department of Physics and Astronomy, Iowa State University, Ames, Iowa 50011, USA}
\author{Mamoon A. Sharaf}
 \email{Corresponding author: msharaf@impcas.ac.cn}
\affiliation{Institute of Modern Physics, Chinese Academy of Sciences, Lanzhou 730000, China}
\affiliation{Department of Physics and Astronomy, Iowa State University, Ames, Iowa 50011, USA}
\author{James P. Vary}
\affiliation{Department of Physics and Astronomy, Iowa State University, Ames, Iowa 50011, USA}

\maketitle

\section{Introduction}

Light-Front Dynamics (LFD), originally proposed by Dirac \cite{Dirac:1949}, provides a powerful framework for solving relativistic quantum field theories (QFTs) by quantizing on a null plane.~The Light-Front Hamiltonian formalism simplifies the vacuum structure and eliminates the need for kinematic boost operators, making it a natural choice for describing bound-state phenomena in high-energy physics, particularly in the context of the parton model \cite{Brodsky:1998,HillerReview:2016}.

A primary non-perturbative numerical tool within LFD is the Discretized Light-Cone Quantization (DLCQ) \cite{Brodsky:1985}, which introduces box-normalized standing waves for each spatial direction.~For the longitudinal direction where total momentum is conserved, it introduces a momentum resolution parameter, $K$, when discretizing the longitudinal momentum 
$P^+$.
Following
its 
initial application to solving two-dimensional 
(2D) interacting scalar 
$\phi^4$ theory \cite{Harindranath1987}, DLCQ has 
been 
hindered by computational limits. 
In particular, for DLCQ studies of 2D $\phi^4$ theory
using Periodic Boundary Conditions (PBCs), the constrained zero momentum mode ($\phi_0$) is 
conventionally
omitted to avoid the 
divergence linked with
its nonlinear operator constraint equation \cite{Maskawa:1976} and to avoid the divergence in the zero-mode kinetic energy.~However, neglecting $\phi_0$ introduces systematic errors, leading to slow convergence of calculated observables toward the continuum limit, generally scaling as $O(1/K)$~\cite{Hiller:2009}.

The 2D interacting scalar field $\phi^{4}$ theory serves as 
a challenging
testbed for these non-perturbative methods, exhibiting a second-order phase transition from a symmetric phase to a spontaneously broken phase \cite{Chang:1976}.~Our previous work \cite{Vary:2022}, which utilized uncorrected
(i.e. no zero-mode correction)
DLCQ with PBCs, successfully determined the critical coupling for this phase transition, achieving consistency with results obtained via light-cone conformal truncation \cite{Anand:2017} (see Table 1 in each of Refs.~\cite{DemirayHTEFT2025, Chabysheva2025_Zero_Modes} for additional critical coupling results using different techniques).  
\color{black}

Crucially, however, attaining this level of accuracy required reaching a very high resolution of $K=88$.~As the complexity of the DLCQ basis states is governed by the partition function, this high cutoff required diagonalizing Hamiltonian matrices with dimensions exceeding 22 million states per sector, placing extensive demands on high-performance computing resources.    

The difference between equal-time and light-front time results for the value of the extrapolated critical coupling was not addressed though an appreciation of the importance of neglected zero-mode contributions had emerged~\cite{Burkhardt2016}.  As discussed in Ref. \cite{Burkhardt2016}, the discrepancy between light-front and equal-time quantization is due to mass renormalization.  In particular, the bare mass in equal-time quantization is perturbatively renormalized by tadpole contributions in equal-time quantization but not in light-front quantization.  Though both quantizations yield the same physical quantities, the situation with critical coupling, where the physical mass scale goes to zero, is more complicated \cite{Vary:2022, Chabysheva2025_Zero_Modes, Vary:2025yqo}. \color{black}

The current work
addresses the long-standing challenge of poor convergence in DLCQ. Building upon theoretical proposals for including zero modes as perturbative corrections \cite{Hiller:2009}, we 
incorporate zero-mode contributions into the effective DLCQ Hamiltonian for 2D $\phi^{4}$ theory.~We apply the correction perturbatively to the eigenvalues of the original DLCQ Hamiltonian to obtain a set of corrected critical coupling values at finite $K$.

We demonstrate that this 
zero-mode
correction dramatically accelerates the convergence of the critical coupling in the continuum limit.~This methodological improvement is quantified by a significant reduction in the required resolution:~comparable accuracy to the uncorrected $K=88$ result is achieved with the corrected data at $K=70$.~This shift dramatically lowers the size of the required basis space, reducing the necessary matrix dimension by an order of magnitude (from approximately 22 million states to 2 million states per sector).~This result validates the use of perturbative zero-mode corrections as a highly efficient mechanism for advancing DLCQ and related Hamiltonian Light-Front techniques, such as Basis Light-Front Quantization (BLFQ) \cite{Vary:2010}, where computational efficiency is paramount for 
studies in $3+1$ dimensions.

In the higher-dimensional context, BLFQ serves as a framework for investigating hadronic structure, including the mass spectroscopy of heavy quarkonia, the extraction of generalized parton distributions, and the calculation of electromagnetic form factors for nucleons and mesons \cite{Mondal:2020,Xu:2024sjt,Vary:2025yqo}. 
The acceleration of convergence demonstrated here suggests a potentially useful pathway to treat zero-mode corrections through effective interactions for these computationally intensive studies to attain the high-precision regime required for comparison with modern experimental data.

To be more concrete, the implementation of zero-mode corrections in this work supports the further development of non-perturbative Hamiltonian light-front methods 
that even more completely account for zero-mode contributions
as discussed Fitzpatrick, Katz, and Walters \cite{Fitzpatrick2020} and proposed by Chabysheva and Hiller \cite{Chabysheva2025_Zero_Modes}. \color{black}  
 Our results demonstrate that 
an
effective Hamiltonian 
with
zero-mode contributions 
can be a suitable substitute for 
explicit inclusion of the $K^+=0$ mode in the basis space.~By involving these theoretical corrections directly in the effective Hamiltonian, one retains the ability to work within a
conventional
DLCQ framework while capturing 
vacuum effects usually lost when the zero mode is omitted.~This approach 
suggests a
pathway for higher-dimensional gauge theories, where the $K^+=0$ mode often introduces intractable divergences.~Specifically, the heuristic insights gained from this scalar $\phi^4$ study regarding conservation rules 
may prove generalizable to the gluon sector in 3+1 
dimensional QCD.
\section{Formalism and zero-mode correction}

\subsection{DLCQ Hamiltonian}

\hspace{2.5mm} The 2D $\phi^4$ theory in light-cone quantization with PBCs and neglecting the constrained zero mode \cite{Harindranath1987} is defined by the dimensionless mass-squared operator $M^2  = KH$, where $K$ is the total dimensionless longitudinal momentum operator given in normal order by
\begin{equation}
K=\sum_{n}na^{\dagger}_n a_{n},
\label{eqK}
\end{equation}
where $a_n^{\dagger}$ and $a_n$ are creation and annihilation operators for bosons with dimensionless integer longitudinal momentum $n \ge 1$.  The normal-ordered DLCQ Hamiltonian $H$ is given \mbox{by:}
\begin{multline}
H = \mu^{2} \sum_{n} \frac{1}{n}\, a^{\dagger}_n a_n + \frac{\lambda}{16\pi} \sum_{k,l,m,n} \frac{a^{\dagger}_k a^{\dagger}_l a_{m} a_{n}}{\sqrt{k l m n}}\, \delta_{m+n,k+l} \\
 + \frac{\lambda}{24\pi} \sum_{k,l,m,n} \frac{a^{\dagger}_k a_{l} a_{m} a_{n} + a^{\dagger}_k a^{\dagger}_l a^{\dagger}_{m} a_{n}}{\sqrt{k l m n}}\, \delta_{m+n,k+l},
\end{multline}
where $\lambda$ is the coupling constant.~The parameter $\mu^2$ is the bare-mass squared of the scalar field which we choose to be equal to 1 throughout.  For this reason, our results can be interpreted as results for $\lambda/\mu^{2}$. \color{black}

We construct the dimensionless mass-squared operator $M^{2}=KH$ by utilizing a general state in the Fock space basis denoted as $|n_{1}^{m_{1}}, n_{2}^{m_{2}},n_{3}^{m_{3}},...\rangle$ in order to represent $m_{1}$ quanta in $n_{1}$ units of momentum and so on.~Then, for given $K$, one has $K=n_{1}m_{1}+n_{2}m_{2}+...~$.~In this case, $K$ is the constraint upon which basis states are formed.~So, the value of $K$ dictates the dimension of the Hamiltonian matrix.~Given $K$,~basis states are partitions of $K$ and therefore, the dimensions of the Hamiltonian matrix will grow rapidly with $K$.  We note that the Hamiltonian ($H$) relates to the longitudinal momentum ($K$) via $M^2 = KH = P^+P^-$.  We solve $M^2=KH$ to obtain the low-lying dimensionless eigenvalues $M^2(K,\lambda/\mu^2)$ and their eigenvectors.  See Refs.~\cite{Harindranath1987, Vary:2022, ShreeramThesis} for details.    \color{black}
\newline

\subsection{Systematic zero-mode correction}
\hspace{2.5mm} As mentioned above, the exclusion of the zero-mode $\phi_0$ introduces systematic truncation errors that manifest as slow convergence, scaling as $O(1/K)$, which necessitated calculations up to $K=88$ to attain a reasonable estimate of the continuum limit $K \rightarrow \infty$~\cite{Vary:2022}.~To systematically improve this convergence, we now incorporate the zero-mode field into the effective Hamiltonian perturbatively.


The constrained nature of the zero mode allows its contributions to be derived as an effective interaction, $H^{\text{zero mode}}$, involving only the dynamical modes as shown in Ref. \cite{Hiller:2009}. The resulting normal-ordered expression for the zero-mode correction to the Hamiltonian is given by:

\begin{widetext}
\begin{equation}
\begin{aligned}
H^{\text{zero mode}} ={}&
 -\frac{2\mu^2 g^2}{4}
  \sum_{n,m,l,p,q,s>0}
  \frac{\delta_{n+m-l}\,\delta_{p+q-s}}{\sqrt{n m l p q s}}
  \Bigl(
    2 a_n^\dagger a_m^\dagger a_s^\dagger a_p a_q a_l
    + a_n^\dagger a_m^\dagger a_p^\dagger a_q^\dagger a_l a_s
    + a_l^\dagger a_s^\dagger a_n a_m a_p a_q
  \Bigr)
\\[0.4em]
& -\frac{2\mu^2 g^2}{2}
  \sum_{n,m,p,q>0}
  \frac{1}{n+m}\,
  \frac{\delta_{n+m+p-q}}{\sqrt{n m p q}}
  \Bigl(
    a_n^\dagger a_m^\dagger a_p^\dagger a_q
    + a_q^\dagger a_n a_m a_p
  \Bigr)
\\[0.4em]
& -\frac{2\mu^2 g^2}{4}
  \sum_{n,m,p,q>0}
  \left(
    \frac{1}{n+m}
    + \frac{4\,\Theta(q-n)}{q-n}
  \right)
  \frac{\delta_{n+m-p-q}}{\sqrt{n m p q}}\,
  a_n^\dagger a_m^\dagger a_p a_q
\\[0.4em]
& -\frac{2\mu^2 g^2}{6}
  \sum_{m>n>0}
  \frac{1}{m n(m-n)}
  \Bigl(
    a_m^\dagger a_m
    + 4 a_n^\dagger a_n
  \Bigr),
\end{aligned}
\label{eqZeroModeHamiltonian}
\end{equation}
\end{widetext}
where $\Theta(x)$ is the Heaviside function and $g = \frac{\lambda}{4 \pi \mu^2} $ so that the zero-mode correction can be expressed compactly in terms of $g^2$.

We note that in general, incorporating the zero mode in DLCQ is a nontrivial problem, as discussed in Refs.~\cite{Vary:2022,mengyao_masters_thesis}.~This is because the zero-mode operator is constrained and obeys a nonlinear operator equation.~
In typical DLCQ calculations, zero modes are neglected because of the complexity introduced in solving the associated constraint equations, which could lead to numerical
difficulties.
~In our analysis, we found that simply adding 
the
zero-mode Hamiltonian contribution 
of Eq.~\eqref{eqZeroModeHamiltonian}
to the regular Hamiltonian did not produce results consistent with the expected trend.~Instead, the calculated observables exhibited irregular or inconsistent behavior, indicating that a naive additive treatment of the zero mode is insufficient and that its constrained dynamics must be treated more carefully.


The terms in this expression represent effective interactions originating from the zero-mode field exchange and require appropriate regularization to handle sums that diverge in the continuum limit.~See Section \ref{SecS1} in Supplemental Material for a demonstration of how the zero-mode matrix elements are calculated for $K=6$ and Ref. \cite{ShreeramThesis} for more details.

The corrected lowest mass-squared eigenvalue, $M^2_{\text{corrected}}$, at finite resolution $K$ is then obtained via a first-order perturbative correction:
\begin{equation}
M^2_{\text{corrected}} = 
M^2 + \langle \Phi_{0} | KH^{\text {zero mode }} | \Phi_{0} \rangle,
\end{equation}
where $| \Phi_{0} \rangle$ is the lowest eigenvector of the uncorrected Hamiltonian $H$ and $M^2$ is its corresponding eigenvalue. By using this corrected eigenvalue, we can determine a new set of critical coupling values, $\lambda_{c, \textrm{corrected}}(K)$, which are expected to converge more rapidly to the continuum limit.


\section{Results and discussion}

\subsection{Accelerated convergence in the odd sector}

\begin{figure}[t]
  \centering
  \includegraphics[width=\columnwidth]{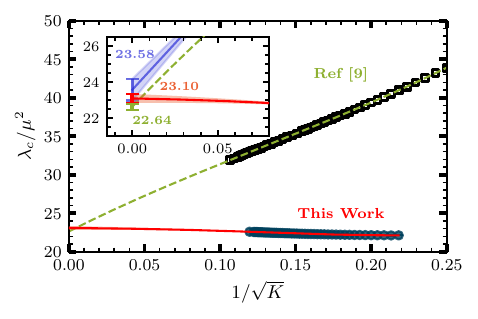}
  \caption{Critical coupling $\lambda_c/\mu^{2}$ in 2D $\phi^4$ theory as a function of $1/\sqrt{K}$. Our new results (filled blue circles) are compared against the uncorrected data from Ref.~\cite{Vary:2022} (hollow black squares).~The red solid line represents the Gaussian Process Regression (GPR, see Section~\ref{SecIV}) mean prediction for the zero-mode corrected data (blue circles), with the red shaded region indicating the $1\sigma$ uncertainty band. The resulting GPR limit 
  (red line's intercept)
  is $\lambda_c/\mu^{2} = 23.10 \pm 0.25$.~The inset displays a detailed comparison of extrapolation methods for the Ref.~\cite{Vary:2022} dataset:~the original 5th-degree polynomial fit (green dashed line) yielding $\lambda_c/\mu^{2} = 22.64 \pm 0.17$, and our GPR re-analysis of the same data (blue solid line) yielding $\lambda_c/\mu^{2} = 23.58 \pm 0.60$.}
    \label{fig:GPR_extrapolation}
\end{figure}

\hspace{2.5mm} We apply the perturbative zero-mode correction $H^{\text {zero mode }}$ to the lowest state of the odd sector, which determines the critical coupling $\lambda_c/\mu^{2}$ for the vanishing mass gap.~Our primary result is the dramatic acceleration of convergence toward the continuum limit.~In Fig.~\ref{fig:GPR_extrapolation}, we compare the discrete critical coupling values obtained from the original uncorrected DLCQ calculation \cite{Vary:2022} with the new, corrected values, plotted as a function of the resolution parameter $1/\sqrt{K}$.

The black hollow data points denote the original (uncorrected) $\lambda_{c}/\mu^{2}$ data~\cite{Vary:2022} as a function of $1/\sqrt{K}$ along with 
the polynomial fit (green dashed line) of Ref.~\cite{Vary:2022}.
In contrast, the blue data points  denote the (corrected) $\lambda_{c}/\mu^{2}$ data of the present work obtained using $H^{\textrm{zero mode}}$.  This visual comparison confirms that the zero-mode correction successfully mitigates the dominant systematic truncation errors, consistent with the improved convergence analyzed by Chabysheva and Hiller 
\cite{Hiller:2009}. The various lines denote results of fitting processes  described below.

Across the full range of $K$ studied, the zero-mode corrected critical couplings lie much closer to the eventual continuum limit than the corresponding uncorrected values. In particular, at $K$=60 the uncorrected coupling $\lambda_c/\mu^2 \approx 33.7$, is still far from the continuum limit, whereas the corresponding corrected coupling \mbox{$\lambda_c/ \mu^2 \approx 22.5$} is very close to the eventual continuum limit.

Our findings are consistent with the conclusions of Ref.~\cite{Hiller:2009}, who argued that including the constrained zero mode as an effective interaction is expected to suppress the leading $O(1/K)$ truncation errors of DLCQ and improve the convergence of the ground state eigenvalues \color{black} to $O(1/K^2)$.  See Fig. \ref{fig:GSEigenvalue} in Section \ref{SecS2} of Supplemental Material, which shows convergence of the ground state eigenvalue with $1/K$. \color{black}

Taken together with our earlier analysis \cite{Vary:2022} and related continuum approaches \cite{Anand:2017}, these observations show that the zero-mode corrected critical couplings enter the expected continuum window already at moderate $K$, whereas the uncorrected values remain significantly biased even at the highest resolutions we can currently attain. See Table \ref{tab:critical_coupling} in Section \ref{SecS3} of Supplemental Material for a numerical comparison of the values of the uncorrected critical coupling ($\lambda_{c, \textrm{orig}}/\mu^{2}$) from Ref.~\cite{Vary:2022} and the corrected ($\lambda_{c, \textrm{corr}}/\mu^{2}$) at different $K$.

\subsection{Computational efficiency and resource reduction}

\hspace{2.5mm} The primary benefit of accelerated convergence is the reduction in required computational resources.~In our previous work, robust extrapolation required calculating data points up to a resolution of $K_{\text{max}} = 88$.~Using the zero-mode corrected formalism, the critical couplings achieve comparable stability but with remarkable proximity to the continuum limit at a significantly reduced maximum resolution of $K_{\text{max}} = 70$.

This might seem to be a small reduction in $K$ but it yields an exponential decrease in the size of the required basis space, which is governed by the partition function:
\begin{itemize}
\item $K=88$ (upper limit of uncorrected calculations): 
DLCQ basis dimension is approximately 22 million states for each (even/odd) sector.
\item $K=70$ (upper limit of corrected calculations):
DLCQ basis dimension is approximately 2 million states for each (even/odd) sector.
\end{itemize}

This order-of-magnitude reduction in the Hamiltonian matrix dimension to attain comparable accuracy translates into an approximate factor of 100 reduction in both memory and CPU time requirements.
%


\section{Gaussian process regression for continuum extrapolation}
\label{SecIV}

\subsection{Continuum extrapolation using Gaussian process regression}

\hspace{2.5mm} To minimize model dependence and avoid the unphysical artifacts often associated with high-degree polynomial fits, we employ Gaussian Process Regression (GPR) to perform the meta-extrapolation of our windowed intercepts.~Unlike parametric methods that force data into a global functional form, GPR is a non-parametric Bayesian approach that treats the fit as a distribution over possible functions~\cite{Rasmussen2006,Bijnens2019}.

We utilize a Matern kernel (see Section~4 in Ref.~\cite{Rasmussen2006}) with smoothness parameter $\nu=1.5$ to balance local flexibility with global trend stability. This kernel is particularly effective for DLCQ data, as it efficiently handles the increased level density and numerical noise encountered near the vanishing mass gap.~The resulting $1\sigma$ and $2\sigma$ uncertainty bands provide a rigorous, Bayesian-derived estimate of the predictive variance as $1/\sqrt{K}\rightarrow 0$.

\begin{figure}[h]
    \centering
    \includegraphics[width=\columnwidth]{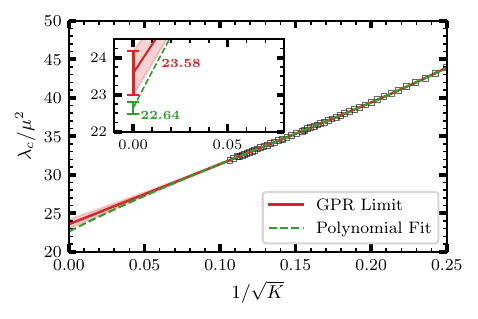}
    \caption{GPR continuum extrapolation applied to the uncorrected 
    DLCQ data of Ref.~\cite{Vary:2022}, yielding 
    $\lambda_c/\mu^{2} = 23.58 \pm 0.60$. The shaded band indicates the $1\sigma$  predictive uncertainty.~This result is 
    consistent with the zero-mode corrected value of 
    $23.10 \pm 0.25$ obtained in the present work.}
    \label{fig:gpr_old}
\end{figure}

Notably, when this GPR framework is applied to the benchmark data from Ref.~\cite{Vary:2022}, we find a continuum limit of $\lambda_c/\mu^{2}=23.58\pm 0.60$ which is somewhat different from the originally published polynomial-based extrapolation of $\lambda_c/\mu^{2}=22.64\pm 0.17$ as seen in Fig.~\ref{fig:gpr_old}.  Our current GPR reanalysis of the original benchmark data is, however, consistent with our new result, using the perturbative zero-mode correction, of $23.10\pm0.25$, suggesting that GPR effectively identifies a stable plateau in the coupling that traditional polynomials may miss due to sharp convergence behaviors at small $1/\sqrt{K}$.

A log-log regression of the residuals $|(\lambda_c(K) - \lambda_c^{\text{cont}})/\mu^{2}|$ 
yields an effective scaling exponent of $\alpha \approx 0.53$ for the 
uncorrected data
and $\alpha \approx 0.57$ \footnote{Note that the correct value of the critical exponent is $\alpha=0.125$.  This discrepancy is due to the need for a fully non-perturbative treatment and due to additional zero-mode corrections that are absent from $H^{\textrm{zero mode}}$.} for the perturbatively corrected data,  indicating that $\lambda_c/\mu^{2}$ converges 
approximately as $1/\sqrt{K}$ rather than as the $O(1/K)$ predicted for 
individual matrix elements~\cite{Hiller:2009}\color{black}. This reflects the fact that 
$\lambda_c/\mu^{2}$ is a nonlinear functional of a non-perturbative procedure, and its 
effective convergence rate need not match that of the underlying Hamiltonian 
truncation.~Crucially, the zero-mode correction does not alter the scaling 
exponent but substantially reduces the pre-factor, shifting the corrected 
data much closer to the continuum value at every $K$. This behavior — 
improved intercept rather than improved power law — is precisely the regime 
where GPR excels over parametric polynomial or power-law fits, as it 
captures the global convergence trend without assuming a specific functional 
form.

\subsection{Stability of GPR continuum extrapolation}
\hspace{2.5mm} While the Matérn GPR framework is non-parametric, it is important to 
confirm that the continuum extrapolation is not dominated by the 
highest-resolution data points nearest the intercept. To test this, 
we iteratively remove the largest-$K$ points from the fit range and 
monitor the resulting shift in the GPR meta-limit $\lambda_c/\mu^{2}$.

As shown in Tables~\ref{tab:low_res_stability} and~\ref{tab:high_res_stability}, the GPR limit remains 
remarkably stable in both datasets as the lowest and highest-resolution points 
are removed
respectively,
shifting by less than $0.5\%$ across all truncations.~To better display this stability, we present our data in tables with 3 significant figures beyond the decimal point.
This stability confirms that the GPR successfully captures the global 
convergence trend rather than overfitting to the points nearest the 
continuum intercept, and that the extrapolated value of $\lambda_c/\mu^{2}$ 
is robust against the precise choice of $K_{\text{max}}$.
\begin{table}[t] 
\centering
\caption{Unified low-resolution ($K_{\min}$) stability analysis for the odd sector with fixed $K_{\max}$(70 for current data and 88 for old data). The continuum limit $\lambda_c/\mu^{2}$ is compared as a function of the minimum resolution floor. Dashes indicate resolution thresholds where data points are not present in the respective dataset, precluding a GPR extrapolation.}
\label{tab:low_res_stability}
\begin{tabular}{ccc}
\hline\hline
$K_{\min}$ &  $\lambda_c/\mu^{2} \pm \sigma$ & Ref  \cite{Vary:2022} \\
\hline
12 & ---                & $23.577 \pm 0.598$ \\
16 & ---                & $23.557 \pm 0.459$ \\
20 & ---                & $23.544 \pm 0.371$ \\
21 & $23.096 \pm 0.247$ & $23.541 \pm 0.354$ \\
22 & $23.120 \pm 0.249$ & $23.539 \pm 0.339$ \\
23 & $23.109 \pm 0.252$ & $23.537 \pm 0.324$ \\
24 & $23.108 \pm 0.255$ & $23.536 \pm 0.311$ \\
25 & $23.111 \pm 0.257$ & $23.534 \pm 0.299$ \\
26 & $23.111 \pm 0.260$ & $23.533 \pm 0.288$ \\
27 & $23.095 \pm 0.263$ & $23.532 \pm 0.278$ \\
28 & $23.105 \pm 0.266$ & $23.531 \pm 0.269$ \\
29 & $23.092 \pm 0.269$ & $23.531 \pm 0.262$ \\
30 & $23.086 \pm 0.272$ & $23.531 \pm 0.260$ \\
\hline\hline
\end{tabular}
\end{table}

\begin{table}[t]
\centering
\caption{Unified high-resolution ($K_{\max}$) stability analysis for the odd sector with fixed $K_{\min}$(21 for current data and 12 for old data). The continuum limit $\lambda_c/\mu^{2}$ is monitored as the high-resolution boundary of the data window is systematically lowered.~Dashes indicate where data points are unavailable due to the specific $K$-sampling of each study.}
\label{tab:high_res_stability}
\begin{tabular}{ccc}
\hline\hline
$K_{\max}$ &  $\lambda_c/\mu^{2} \pm \sigma$ & Ref. \cite{Vary:2022} \\
\hline
60 & $22.909 \pm 0.249$ & $23.702 \pm 0.942$ \\
62 & $22.954 \pm 0.249$ & $23.698 \pm 0.914$ \\
63 & $22.977 \pm 0.250$ & --- \\
64 & $22.998 \pm 0.249$ & $23.693 \pm 0.887$ \\
65 & $23.019 \pm 0.249$ & --- \\
66 & $23.039 \pm 0.249$ & $23.688 \pm 0.863$ \\
67 & $23.058 \pm 0.249$ & --- \\
70 & $23.096 \pm 0.247$ & $23.675 \pm 0.818$ \\
72 & ---                & $23.668 \pm 0.797$ \\
74 & ---                & $23.660 \pm 0.778$ \\
76 & ---                & $23.652 \pm 0.759$ \\
80 & ---                & $23.635 \pm 0.725$ \\
84 & ---                & $23.621 \pm 0.699$ \\
88 & ---                & $23.604 \pm 0.671$ \\
\hline\hline
\end{tabular}
\end{table}
To assess the robustness of our results against potential non-asymptotic behavior at the boundaries, we perform a symmetric pruning analysis by iteratively narrowing the fit window
as presented in Table~\ref{tab:symmetric_pruning}.
~This systematic removal of both the lowest and highest resolution points ensures that the extrapolated continuum limit is anchored by the dense, high-fidelity bulk of the dataset rather than edge-case numerical artifacts.
\begin{table}[H]
\centering
\caption{Symmetric pruning stability analysis for the odd sector. The continuum limit $\lambda_c/\mu^{2}$ is monitored as the data window is narrowed symmetrically from both the low-resolution ($K_{\min}$) and high-resolution ($K_{\max}$) boundaries.~The consistent behavior of the intercept across 10 pruning steps demonstrates that the GPR limit is anchored by the high-density bulk of the dataset.~Dashes indicate where data points are unavailable due to specific $K$-sampling of each study.~$N$ is the number of data points in each $K$ range chosen.}
\label{tab:symmetric_pruning}
\begin{tabular}{cccccc}
\hline\hline
\multirow{2}{*}{$K$ Range} & \multicolumn{2}{c}{This Work} & & \multicolumn{2}{c}{Ref. \cite{Vary:2022}} \\
\cline{2-3} \cline{5-6}
& $\lambda_c/\mu^{2} \pm \sigma$ & $N$ & & $\lambda_c/\mu^{2} \pm \sigma$ & $N$ \\
\hline
$[12, 88]$ & --- & -- & & $23.604 \pm 0.671$ & 52 \\
$[16, 80]$ & --- & -- & & $23.599 \pm 0.545$ & 46 \\
$[20, 74]$ & --- & -- & & $23.595 \pm 0.473$ & 39 \\
$[21, 70]$ & $23.096 \pm 0.247$ & 48 & & $23.600 \pm 0.480$ & 36 \\
$[22, 66]$ & $23.064 \pm 0.251$ & 45 & & $23.604 \pm 0.494$ & 33 \\
$[23, 64]$ & $23.011 \pm 0.255$ & 42 & & $23.602 \pm 0.495$ & 31 \\
$[24, 62]$ & $22.964 \pm 0.258$ & 39 & & $23.601 \pm 0.501$ & 29 \\
$[25, 60]$ & $22.919 \pm 0.262$ & 36 & & $23.600 \pm 0.513$ & 27 \\
$[26, 58]$ & $22.864 \pm 0.264$ & 33 & & $23.599 \pm 0.533$ & 25 \\
$[30, 56]$ & $22.734 \pm 0.267$ & 27 & & $23.612 \pm 0.595$ & 20 \\
\hline\hline
\end{tabular}
\end{table}

The symmetric pruning results presented in Table~\ref{tab:symmetric_pruning} provide the most rigorous check of the GPR framework.~By narrowing the data window from both ends, we demonstrate that the current work maintains a stable intercept even when the basis size is reduced from $N=48$ to $N=27$. Notably, the statistical uncertainty in our most heavily pruned window ([30,56]) remains significantly lower ($\sigma$=0.267) than the uncertainty of the full Ref.~\cite{Vary:2022} dataset ($\sigma$=0.671). This more than two-fold improvement in precision, despite a smaller total number of points, highlights the efficiency gains afforded by the zero-mode corrected data.~Furthermore, while the Ref.~\cite{Vary:2022} dataset exhibits increased fluctuations in $\lambda_c/\mu^{2}$ as points are removed, our results show a smooth, monotonic convergence, suggesting that the corrected data successfully eliminates the non-asymptotic chatter typically seen in low-resolution DLCQ calculations.

Our 
even-sector
results are consistent with those of the 
odd
sector. However, a quantitative GPR extrapolation 
for the even sector
analogous to that of the odd sector is not appropriate. 
See Section \ref{SecS4} in Supplemental Material for more details, including Figs.~\ref{fig:evenzeromode} and \ref{fig:4}.

\subsection{Kernel sensitivity}

\hspace{2.5mm} To assess the sensitivity of the continuum extrapolation to the 
choice of kernel smoothness, we evaluated the log marginal likelihood 
(LML) across the standard Mat\'{e}rn half-integer values.~The results 
--- \mbox{$\nu = 0.5$} (LML: 27.6), $\nu = 1.5$ (LML: 48.9), and $\nu = 
2.5$ (LML: 51.2) --- show a dramatic improvement from $\nu = 0.5$ 
to \mbox{$\nu = 1.5$}, confirming that the convergence data is genuinely 
smooth rather than merely continuous.~The large jump of $\Delta
\text{LML} \approx 21$ between $\nu = 0.5$ and \mbox{$\nu = 1.5$} indicates 
that once-differentiable sample paths are strongly preferred by the 
data over the rougher Ornstein-Uhlenbeck kernel.

While $\nu = 2.5$ achieves a marginally higher LML score, the 
difference of $\Delta\text{LML} \approx 2.3$ relative to $\nu = 1.5$ 
is modest and does not constitute decisive evidence for 
twice-differentiable sample paths. We therefore adopt $\nu = 1.5$ 
as the minimal smoothness assumption consistent with a well-defined, 
once-differentiable continuum limit, avoiding the imposition of 
higher regularity than is physically warranted by the finite-$K$ 
discretization structure of DLCQ data.

As a direct numerical validation, fitting with $\nu = 2.5$ yields 
$\lambda_c/\mu^{2} = 23.21$, consistent with the $\nu = 1.5$ result of 
$\lambda_c/\mu^{2} = 23.10 \pm 0.25$ within $1\sigma$.~Combined with the 
data truncation stability demonstrated in 
Tables~\ref{tab:low_res_stability},~\ref{tab:high_res_stability} and ~\ref{tab:symmetric_pruning}, this confirms that the continuum estimate 
is robust with respect to both the choice of kernel smoothness and 
the precise set of $K$ values included in the fit.

\section{Conclusion}

This work demonstrates a major advancement in the application of DLCQ to non-perturbative quantum field theories.~While our previous effort successfully established the critical coupling $\lambda_c/\mu^{2}$ for $2D$ $\phi^4$ theory (omitting the zero mode) by extrapolating extensive finite-$K$ data calculated up to $K=88$, this achievement came at an immense computational cost due to the slow $O(1/K)$ convergence.

By systematically 
incorporating
the analytically derived zero-mode correction $H^{\text{zero mode}}$ as a perturbative correction to the DLCQ mass eigenvalues, we have achieved a significant and decisive acceleration in convergence. This methodological improvement has two immediate and important consequences:

\begin{enumerate}
\item Validation and consistency:~The zero-mode corrected critical couplings for both the odd and even sectors converge rapidly, confirming the physical consistency of the DLCQ approach when zero-mode effects are systematically included.~The clustering of corrected data around the established continuum value even at low $K$ removes the finite-resolution divergence observed in the uncorrected data.
\item Computational efficiency:~The corrected formalism achieves a stable result in the critical region at a maximal resolution of $K_{\text{max}} = 70$, down from $K_{\text{max}} = 88$.~This reduction corresponds to an order-of-magnitude decrease in required basis states,~from approximately 22 million to 2 million per sector (odd or even). This dramatic computational savings validates the inclusion of zero-mode corrections as a necessary and practical tool for resource-efficient high-precision DLCQ calculations. 
\end{enumerate}

The successful implementation of $H^{\text {zero mode }}$ provides a clear pathway for addressing systematic errors in other complex light-front Hamiltonian applications.~This is particularly relevant for the BLFQ program applied to realistic theories in $3+1$ dimensions, where mitigating the exponential growth of basis states is a primary challenge to achieving systematic continuum extrapolations.

Based on the established relationships between the odd and even sectors in this theory \cite{Vary:2022}, we find that the perturbative corrections to the even sector yield critical couplings 
at fixed $K$
consistently lower than uncorrected results, mirroring the behavior observed in the odd sector. While the even-sector data is subject to reduced density due to the partitioning of even and odd $K$ branches, the downward shift in $\lambda_c/\mu^{2}$ provides sufficient confirmation of cross-sector consistency.~Consequently, we focus our detailed stability analysis on the higher-density odd-sector results and do not explicitly display the even-sector curves here.

Overall, incorporating the analytically derived zero-mode interaction into DLCQ provides a practical way to suppress leading discretization errors in $\lambda_c/\mu^{2}$ while reducing the required basis size by roughly an order of 
magnitude. We believe that a similar
strategy can be extended to more complex light-front Hamiltonian frameworks such as
the application of BLFQ to gluon dynamics in QCD.

\makeatletter
\newcommand{\mysection}[1]{%
  \par\vspace{2.5ex}%
  \begin{center}
    \normalfont\small\bfseries #1
  \end{center}
  \vspace{1.5ex}
}
\makeatother

\begin{center}
 \mysection{ \normalfont\small \bfseries Declaration of generative AI and AI-assisted technologies in the writing process}
\end{center}
\hspace{2.5mm} During the preparation of this work, the authors used ChatGPT (OpenAI) and Gemini AI (Google) in order to improve language, formatting, and readability.  After using these tools/services, the authors reviewed and edited the content as needed and take full responsibility for the content of the publication.

\begin{center}
 \mysection{ \normalfont\small \bfseries Declaration of competing interest}
\end{center}
\hspace{2.5mm} The authors declare that they have no known competing financial interests or personal relationships that could have appeared to influence the work reported in this paper.

\begin{center}
 \mysection{ \normalfont\small \bfseries Acknowledgments}
\end{center}
\hspace{2.5mm} We are grateful to Dr.~Mengyao Huang for many helpful discussions and contributions during the early stages of this project 
and in the derivation of the zero-mode Hamiltonian.
We also thank Pieter Maris for useful discussions.~This work was supported by US DOE Grants DE-SC0023692 and DE-SC0023707 under the Office of Nuclear Physics Quantum Horizons program for the \textbf{Nu}clei and \textbf{Ha}drons with \textbf{Q}uantum computers (\textbf{NuHaQ}) project.~This research used resources of the National Energy
Research Scientific Computing Center (NERSC), a U.S. Department of Energy Office of Science User Facility located at Lawrence Berkeley National Laboratory, operated under Contract No.~DE-AC02-05CH11231 using NERSC award NP-ERCAP0028672. 
M. A. Sharaf is supported by new faculty startup funding by the Institute of Modern Physics, Chinese Academy of Sciences, grant No.~E539951SBH, by the Gansu International Collaboration and Talents Recruitment Base of Particle Physics (2023–2027), by the Senior Scientist Program funded by Gansu Province grant No.~25RCKA008, and by the Gansu ``Special Program for the Recruitment of Foreign Experts" under grant No.~26RCKA016.\color{black}

 \mysection{ \normalfont\small \bfseries Data availability}
 \hspace{2.5mm} Data will be made available on request.

\title{Critical coupling with zero-mode corrections\\
in discretized light-cone quantized $\phi^4$ Theory}
\date{\today}

\clearpage
\onecolumngrid

\section*{\NoCaseChange{Supplemental material for the paper} 
\\[3ex]
\NoCaseChange{Critical coupling with zero-mode corrections
in discretized light-cone quantized $\phi^4$ Theory by Shreeram Jawadekar, Mamoon A. Sharaf, and James P. Vary}}
\twocolumngrid

\renewcommand\thesection{S\arabic{section}}
\setcounter{section}{0} 

\renewcommand{\thefigure}{S\arabic{figure}}
\setcounter{figure}{0}

\section{Demonstration of zero-mode Hamiltonian calculation}
\label{SecS1}
We demonstrate how elements of $H^{\textrm{zero mode}}$ given by Eq.~\eqref{eqZeroModeHamiltonian} are calculated.  See Ref.~\cite{ShreeramThesis} for additional discussions.  We evaluate the terms:
\begin{itemize}
\item $\sum_{n, m, l, p, q, s>0} \frac{\delta_{n+m-l} \delta_{p+q-s}}{\sqrt{n m l p q s}}(2 a_n^\dagger a_m^\dagger a_s^\dagger a_p a_q a_l )$
\item $\sum_{n, m, l, p, q, s>0} \frac{\delta_{n+m-l} \delta_{p+q-s}}{\sqrt{n m l p q s}} a_n^\dagger a_m^\dagger a_q^\dagger a_s^\dagger a_l a_p $
\item $\frac{g^2}{6} \sum_{m>n>0} \frac{1}{m n(m-n)}\left(a_m^\dagger a_m+4 a_n^\dagger a_n\right)$
\end{itemize}
for $K=6$, where we have five states in the odd sector, namely $\ket{6^1}$,$\ket{4^1,1^2}$,$\ket{3^1,2^1,1^1}$,$\ket{2^3}$,$\ket{2^1,1^4}$.  

For~$\sum_{n, m, l, p, q, s>0} \frac{\delta_{n+m-l} \delta_{p+q-s}}{\sqrt{n m l p q s}}(2 a_n^\dagger a_m^\dagger a_s^\dagger a_p a_q a_l )$, there are both diagonal and non-diagonal matrix elements.  Let us look at the non-diagonal contribution.  The non-diagonal matrix elements give
\begin{equation}
\begin{split}
    &\bra{4^1,1^2} \sum_{n, m, l, p, q, s>0} \frac{\delta_{n+m-l} \delta_{p+q-s}}{\sqrt{n m l p q s}} \\
    &\quad \times (2 a_n^\dagger a_m^\dagger a_s^\dagger a_p a_q a_l ) \ket{3^1,2^1,1^1} 
\end{split}
\tag{S1}
\end{equation}
for the choice $p=3$, $q=1$, $l=2$ and $n=1$, $m=1$, $s=4$.  This term becomes
\begin{equation}
\begin{split}
&\bra{4^1,1^2} \hspace{-6mm} \sum_{n, m, l, p, q, s>0} \frac{\delta_{n+m-l} \delta_{p+q-s}}{\sqrt{n m l p q s}}(2 a_n^\dagger a_m^\dagger a_s^\dagger a_p a_q a_l ) \ket{3^1,2^1,1^1}
\\
 &=2\times 2\frac{\sqrt{1}\times\sqrt{1}\times\sqrt{1}\times \sqrt{1}\times\sqrt{2}\times\sqrt{1}}{\sqrt{4 \times 1\times 1\times 3\times 2\times 1}}.
\end{split}
\tag{S2}
\end{equation}
If $n$ and $m$ were different momenta, then a combinatorial factor of $2$ should be multiplied.  For example, if $p=1$ and $q=3$, then a combinatorial factor of $2$ would be required.

The term $\sum_{n, m, l, p, q, s>0} \frac{\delta_{n+m-l} \delta_{p+q-s}}{\sqrt{n m l p q s}} a_n^\dagger a_m^\dagger a_q^\dagger a_s^\dagger a_l a_p $ is purely a non-diagonal term as it can only increase particle number by 2.  We have:
\begin{equation}
\begin{aligned}
\bra{2^1,1^4}\sum_{n, m, l, p, q, s>0} \frac{\delta_{n+m-l} \delta_{p+q-s}}{\sqrt{n m l p q s}}(a_n^\dagger a_m^\dagger a_q^\dagger a_p^\dagger a_l a_s) \ket{2^3}
\end{aligned}
\tag{S3}
\end{equation}
for $l=2$, $s=2$, $n=m=q=p=1$ which evaluates to
\begin{equation}
\begin{split}
&\bra{2^1,1^4} \sum_{n, m, l, p, q, s>0} \frac{\delta_{n+m-l} \delta_{p+q-s}}{\sqrt{nm l p q s}} (a_n^\dagger a_m^\dagger a_q^\dagger a_p^\dagger a_l a_s) \ket{2^3} \\
&\quad = 2 \frac{\sqrt{3} \times \sqrt{2} \times \sqrt{1} \times \sqrt{2} \times \sqrt{3} \times \sqrt{4}} {\sqrt{2 \times 2 \times 1 \times 1 \times 1 \times 1}}.
\end{split}
\tag{S4}
\end{equation}

The term $\frac{g^2}{6} \sum_{m>n>0} \frac{1}{m n(m-n)}\left(a_m^\dagger a_m+4 a_n^\dagger a_n\right)$ requires careful consideration for convergence since $m$ in the summation doesn't have an upper bound. We have the following cases:
\begin{itemize}
\item For a given state if $m$ is the momentum that exists in the state, then the summation is finite.
\item If both $m$ and $n$ are from the same state, then those summations are finite.
\item When $n$ is set to momentum from the state, then $m$ can take all momenta from $n+1$ to infinity.
\end{itemize}
For the third case where $n$ is fixed to a momentum present in the state and $m$ runs from $n+1$ to $\infty$, we use the standard identity
\begin{equation}
\sum_{m>n>0} \frac{1}{m n(m-n)}=\frac{\psi^{(0)}(n+1)+\gamma}{n^2},
\tag{S5}
\label{eulermascheroni}
\end{equation}
where $\psi^{(0)}(x)$ is the digamma function defined by the recurrence relation:\\ $\psi^{(0)}(x+1)=\psi^{(0)}(x)+\frac{1}{x}$ and $\psi^{(0)}(1)=-\gamma$ and $\gamma$ is the Euler-Mascheroni constant.  If $n$ is the largest momentum of the state, the result is straightforward.  If $n$ is any other momentum then one needs to subtract appropriate terms.

We consider $\ket{4^1,1^2}$ as an example and write
\begin{equation}
    \begin{aligned}
       \bra{4^1,1^2} \sum_{m>n>0} \frac{1}{m n(m-n)}\left(a_m^\dagger a_m+4 a_n^\dagger a_n\right)\ket{4^1,1^2}
    \end{aligned}
    \tag{S6}
\end{equation}
Here we have two momenta 4 and 1. So let us look at the possibilities $n=1$ and $n=4$.  If $n=1$, we have
\begin{equation}
    \begin{aligned}
       \bra{4^1,1^2} \sum_{m>n>0} \frac{1}{m\times 1(m-1)}\left(a_m^\dagger a_m+4 a_1^\dagger a_1\right)\ket{4^1,1^2}
    \end{aligned}
               \tag{S7}
\end{equation}
and $m$ can take all values from 2 to infinity. So, we can use Eq.~\eqref{eulermascheroni} for values of $m$:
\begin{equation}
\begin{aligned}
\bra{4^1,1^2} \sum_{m>n>0} \frac{1}{m\times 1(m-1)}\left(a_m^\dagger a_m+4 a_1^\dagger a_1\right)\ket{4^1,1^2} 
\\= \frac{1}{4 \times 1(4-1)}+4\times2(\frac{\psi^{(0)}(1+1)+\gamma}{1^2}-\frac{1}{4 \times 1(4-1)}).
\end{aligned}
 \tag{S8}
\end{equation}
When $n=4$, it becomes much simpler, that is simply the first  bracket without multiplying factor of 2.
For $m=4$, we took care of that when we added 2nd term in $n=1$.

\section{Convergence of ground state eigenvalues with $1/K$}
\label{SecS2}

\begin{figure}[H]
\centering
\includegraphics[width=\columnwidth]{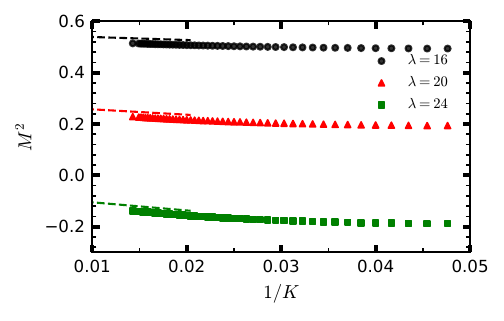}
\caption{Ground state eigenvalue of $M^{2}=KH$ plotted against $1/K$ for different $\lambda$.  Straight line fits to the highest 20 points in $K$ with a small vertical shift are added for visual aid.}
\label{fig:GSEigenvalue}
\end{figure}

Fig.~\ref{fig:GSEigenvalue} shows the corrected ground state eigenvalue of $M^{2}=KH$ plotted against $1/K$ for $\lambda=16, 20, 24$.  Note that as $1/K$ goes to zero, the ground state eigenvalues follow a linear trajectory as evident in the reference straight lines fit to the highest 20 points in $K$.

\section{DLCQ critical coupling values}
\label{SecS3}

To provide quantitative support for the accelerated convergence demonstrated in the main text, we present a selection of the DLCQ critical coupling values for the lowest odd-sector state. As seen in Table~\ref{tab:critical_coupling}, the comparison between the uncorrected value ($\lambda_{c, \text{orig}}$) from our prior work, Ref. \cite{Vary:2022} and the new zero-mode corrected value ($\lambda_{c, \text{corr}}$) at the same resolution $K$ visually confirms the effect of $H^{\text{zero mode}}$ in clustering the data near the continuum limit.

\vspace{0.5em}

\begin{table}[h]
\centering
\caption{Critical coupling comparison for the odd-sector lowest state. The corrected values are significantly closer to the final continuum result than the uncorrected values, even at low $K$.  Column $\lambda_{c, \text{orig}}/\mu^2$ represents Ref.~\cite{Vary:2022} data and $\lambda_{c, \text{corr}}/\mu^2$ represents new corrected data.}
\label{tab:critical_coupling}
\begin{tabular}{|c|c|c|c|}
\hline
$K$ & $1/\sqrt{K}$ & $\lambda_{c, \text{orig}}/\mu^2$ & $\lambda_{c, \text{corr}}/\mu^2$ \ \\
\hline
21 & 0.218 & 40.963 & 22.131 \\
24 & 0.204 & 39.737 & 22.147 \\
27 & 0.192 & 38.750 & 22.178 \\
30 & 0.183 & 37.933 & 22.208 \\
35 & 0.169 & 36.833 & 22.264 \\
40 & 0.158 & 35.959 & 22.320 \\
50 & 0.141 & 34.640 & 22.425 \\
60 & 0.129 & 33.672 & 22.517 \\
\hline
\end{tabular}
\end{table}

\vspace{0.5em}

These data illustrate that while the uncorrected data still require significant extrapolation from \mbox{$\lambda_{c, \text{orig}}/\mu^2 \approx 33.7$} at $K=60$, the corrected data are already clustered tightly around $\lambda_{c, \text{corr}}/\mu^2 \approx 22.5$ in the same finite-$K$ region.

\section{Consistency with even sector}
\label{SecS4}
\begin{figure}[H]
\centering
\includegraphics[width=\columnwidth]{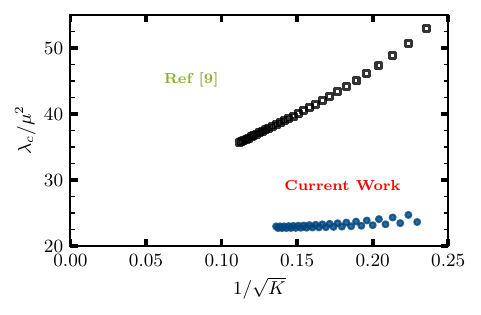}
\caption{Critical coupling $\lambda_c/\mu^2$ for the even sector (lowest state) comparing uncorrected and zero-mode corrected DLCQ data.}
\label{fig:evenzeromode}
\end{figure}

\begin{figure}[h]
    \centering
    \includegraphics[width=\columnwidth]{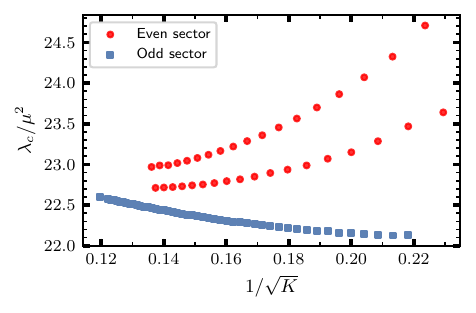}
    \caption{Critical coupling $\lambda_c/\mu^2$ for the lowest states 
    of both the odd sector (blue squares) and even sector (red dots), 
    zero-mode corrected, plotted as a function of $1/\sqrt{K}$. 
    Both sectors converge toward a common continuum value, consistent 
    with the degeneracy condition required for spontaneous symmetry 
    breaking in the continuum limit.}
    \label{fig:4}
\end{figure}

In Figs.~\ref{fig:evenzeromode}
and ~\ref{fig:4}, we demonstrate that the accelerated convergence achieved via the zero-mode correction is not limited to the odd sector. Applying the correction to the lowest state of the even sector yields a similar clustering of corrected critical coupling points near the continuum limit, mitigating the finite-$K$ divergence observed in the uncorrected data. This consistency across both the even and odd sectors further supports the validity and systematic nature of the zero-mode correction, ensuring that the critical coupling extracted still satisfies the necessary degeneracy condition for spontaneous symmetry breaking (SSB) in the continuum limit.

One noticeable feature in the even-sector data is a mild splitting between the critical couplings obtained at even and odd values of $K$, as seen in Fig.~\ref{fig:evenzeromode}.~As the resolution increases, these two sequences move toward one another and approach a common continuum value, indicating that the splitting is a finite-$K$ artifact rather than a physical effect.
\FloatBarrier

While the even-sector corrected data qualitatively converges toward 
the same continuum window as the odd sector (Fig.~\ref{fig:evenzeromode}
and~\ref{fig:4}), a quantitative GPR extrapolation analogous to 
that of the odd sector is not appropriate for this data. The 
alternating even/odd-$K$ splitting present in the even sector 
effectively partitions the data into two interleaved subsequences, 
each with roughly half the number of points available in the odd 
sector. Since GPR is a statistical framework whose reliability depends 
directly on data density, fitting either subsequence independently 
would yield significantly wider uncertainty bands and a poorly 
constrained kernel. The odd-sector GPR result, 
$\lambda_c/\mu^2 = 23.10 \pm 0.25$, therefore serves as the primary 
quantitative continuum estimate of this work. The even-sector data, 
converging visually toward the same continuum window from both 
$K$-parity subsequences, provides independent qualitative support 
for the physical consistency of the zero-mode correction and the 
SSB degeneracy condition in the continuum limit.

\end{document}